\documentclass[aps,physrev,10pt,floatfix,longbibliography,nofootinbib,reprint]{revtex4-2}
\usepackage{amssymb}
\usepackage{bm}
\usepackage{booktabs}
\usepackage{diagbox}
\usepackage{dsfont}
\usepackage{graphicx}
\usepackage[colorlinks,citecolor=blue,linkcolor=blue,urlcolor=blue]{hyperref}
\usepackage{physics}
\usepackage[dvipsnames]{xcolor}

\usepackage[capitalize]{cleveref}

\DeclareMathOperator{\diag}{diag}
\DeclareMathOperator{\Var}{Var}

\begin{document}

\title{Unveiling nonmagnetic phase and many-body entanglement in \\ two-dimensional random quantum magnets Sr$_2$CuTe$_{1-x}$W$_x$O$_6$}
\author{Dian Wu}
\email{dian.wu@epfl.ch}
\affiliation{Institute of Physics, {\'E}cole Polytechnique F{\'e}d{\'e}rale de Lausanne (EPFL), CH-1015 Lausanne, Switzerland}
\author{Fan Yang}
\email{f.yang@epfl.ch}
\affiliation{Institute of Physics, {\'E}cole Polytechnique F{\'e}d{\'e}rale de Lausanne (EPFL), CH-1015 Lausanne, Switzerland}
\author{Giuseppe Carleo}
\email{giuseppe.carleo@epfl.ch}
\affiliation{Institute of Physics, {\'E}cole Polytechnique F{\'e}d{\'e}rale de Lausanne (EPFL), CH-1015 Lausanne, Switzerland}
\date{\today}

\begin{abstract}
We apply a random-plaquette $J_1$-$J_2$ model on the square lattice to capture the physics of a series of spin-$1/2$ Heisenberg antiferromagnet compounds Sr$_2$CuTe$_{1-x}$W$_x$O$_6$.
With the input of experimentally relevant coupling strengths, our exact diagonalization (ED) study probes the ground state properties beyond previous linear spin-wave approach.
An intermediate range of $x \in [0.08, 0.55]$ is identified for a nonmagnetic phase without the long-range N{\'e}el or stripe order. The absence of both valence-bond-glass order and spin-glass non-ergodic dynamics renders its nature intriguing.
Deep inside this phase around $x = 0.3$, we observe signatures potentially linked to randomness-induced short-range spin-liquid-like (SLL) states, including close to zero spin-freezing parameter, vanishing spin-spin correlation beyond nearest neighbors, almost uniform static spin structure factor, as well as a broad tail in the dynamical spin structure factor.
The nonmagnetic phase also features multipartite entanglement in the ground state witnessed by quantum Fisher information (QFI), which exhibits universal scaling behaviors at quantum critical points.
\end{abstract}

\maketitle

\section{Introduction}

Quantum spin liquids (QSLs)~\cite{anderson1987,kivelson1987,laughlin1987}, as ground states of disordered quantum spins characteristic of vanishing long-range magnetic order and massive many-body entanglement, have renewed theoretical and experimental interests~\cite{balents2016,zhou2017,wen2017,knolle2019,li2019,kivelson2020} in recent years.
From a few exactly solvable examples, such as the Kitaev $\mathbb{Z}_2$ spin liquids in the spin-$1/2$ honeycomb model~\cite{kitaev2006} with generalizations to ladder geometries~\cite{feng2007,le2017}, the triangle-honeycomb~\cite{yao2007} and three-dimensional lattices~\cite{maria2016}, as well as extensions to higher-spin analogues~\cite{ma2023}, it is well known that certain types of QSL feature fractional excitations of quasiparticles, including anyons~\cite{wilczek1982} and Majorana zero modes~\cite{majorana1937,alicea2012}, which can be adapted into braiding and fusion mechanisms~\cite{kitaev2006,kee2024} leading to potential fault-tolerant quantum computation~\cite{kitaev2003,nayak2008}.

Pursuing QSLs in experimental platforms proves to be challenging. Quantum simulators~\cite{duan2003,le2018,sagi2019,semeghini2021,nathan2023,sarma2024} offer precise control of spin exchange interactions, yet remain to be scaled up to large system sizes. Quantum materials of frustrated magnets~\cite{balents2010,balents2016,norman2016,li2019,kivelson2020,trebst2022}, on the other hand, deviate from the ideal frustration scenarios set up in simple models, and more often than not involve intrinsic inhomogeneity or quenched randomness, which results in disorder on sites or bonds. Meanwhile, motivated by theoretical concerns on finding more methods to directly observe many-body entanglement as independent evidence for quantum critical phenomena in strongly correlated systems~\cite{song2012,zoller2016,laflorencie2016,irfan2021}, there arise increasing interests and efforts in probing entanglement properties in quantum materials~\cite{le2020,scheie2021,laurell2021,fang2024,ren2024,leiner2024}, and establishing links with quantum information science~\cite{preskill2000,nielsen2010}.

\begin{figure}[t]
\centering
\includegraphics[width=0.8\linewidth]{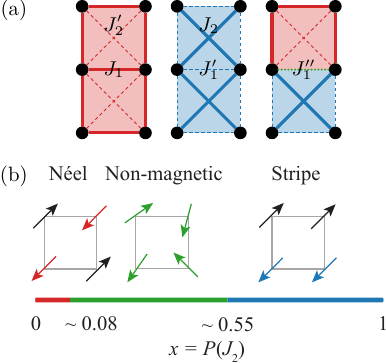}
\caption{
(a) The random-plaquette $J_1$-$J_2$ model of Sr$_2$CuTe$_{1-x}$W$_x$O$_6$. The Heisenberg AFM coupling strengths are obtained from quantum chemistry calculations~\cite{henrik2020e}: $J_1 = 7.4$, $J_2 = 8.3$, $J'_1 = 0.7$, $J'_2 = 0.1$, $J''_1 = 0.3$, in the unit of meV.
Each red (blue) plaquette is dominated by $J_1$ ($J_2$) interactions with a Te (W) ion at the center.
(b) Phase diagram of the ground state via ED as a function of $x$, the mixing ratio of W or the probability of $J_2$-dominated plaquette. At the two ends, two magnetically ordered phases exhibit the N{\'e}el order and the strip order respectively, forming sublattices of spins shown in different colors.
A nonmagnetic phase appears in the middle region $x \in [0.08, 0.55]$.
}
\label{fig:model}
\end{figure}

In this work, we focus on the spin-$1/2$ square-lattice Heisenberg antiferromagnet: the double perovskite Sr$_2$CuTe$_{1-x}$W$_x$O$_6$~\cite{mustonen2018,mustonen2018t,watanabe2018,vasala2014,henrik2016,walker2016,koga2016,henrik2020e,hong2021,henrik2022}, whose degrees of frustration and randomness can be tuned by the mixing ratio $x$ of Te and W ions. At $x = 0$ and $1$, the compounds Sr$_2$CuTeO$_6$ and Sr$_2$CuWO$_6$ are illustrated by the first two panels of \cref{fig:model}~(a). They are dominated by nearest-neighbor $J_1$ and next-nearest-neighbor $J_2$ Heisenberg antiferromagnetic (AFM) interactions and also have weak frustrated interactions $J'_2$ and $J'_1$, thus exhibit the N{\'e}el order and the stripe order respectively, as in \cref{fig:model}~(b).
While nuclear magnetic resonance (NMR) experiments~\cite{mustonen2018,mustonen2018t} suggest a SLL state in between, numerical attempts to reveal its nature become challenging both in the choice of effective spin models, and in the absence of lattice symmetry and translational symmetry due to disorder. Preliminary ED computations on the random-bond $J_1$-$J_2$ model point to a random-singlet phase and a spin-glass phase~\cite{kazuki2018}, while density matrix renormalization group (DMRG) studies find evidence only of the random-singlet state~\cite{ren2023}. The latter is in agreement with quantum Monte Carlo (QMC) simulations on the random $J$-$Q$ model~\cite{liu2018} with $J$ the Heisenberg exchange and $Q$ the multispin interaction. The two effective models studied so far are remotely linked to the material and the role of the mixing ratio $x$ remains obscure.

Here, we focus on a different and more realistic microscopic description~\cite{henrik2020e}, the random-plaquette $J_1$-$J_2$ model depicted in \cref{fig:model}~(a). In the two-dimensional square lattice, each plaquette can be $J_2$-dominated (W counterion) with probability $x$, and otherwise $J_1$-dominated (Te counterion).
When two different kinds of plaquette are adjacent, i.e., when a Te and a W counterions encounter, the strong $J_1$ bond on the shared edge is replaced by a weaker $J''_1$ bond.
From the linear spin-wave approach~\cite{henrik2020e,henrik2022}, the random-plaquette model is found to faithfully reproduce the inelastic neutron scattering spectra, the magnetic excitations, as well as the neutron diffraction patterns of Sr$_2$CuTe$_{1-x}$W$_x$O$_6$. Moreover, Monte Carlo simulations show that even with classical spins, this model can explain the extreme suppression of the N{\'e}el order at small $x = 0.025$, consistent with muon spin resonance ($\mu$SR) measurements~\cite{hong2021}.
Yet, a computational simulation of the quantum model is still lacking.

\begin{figure}[t]
\centering
\includegraphics[width=\linewidth]{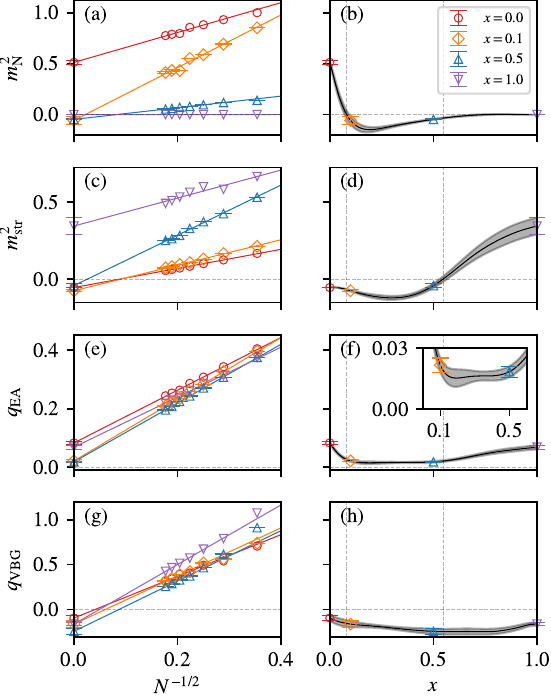}
\caption{
Finite-size scaling of the order parameters $m^2_\text{N}$, $m^2_\text{str}$, $q_\text{EA}$, and $q_\text{VBG}$ used to characterize the phases.
In the left column, for each value of $x$, we measure the order parameters with system sizes $N = 8, 12, 16, 20, 24, 28, 32$ and extrapolate to the thermodynamic limit $N \to \infty$. The error bars are standard errors over random instances and of the extrapolated intercepts.
In the right column, we do the extrapolation for all values of $x$ and plot the results as the black curves. The shaded regions are standard errors of the intercepts. The vertical dashed lines indicate the nonmagnetic phase $x \in [0.08, 0.55]$.
In (b), when $m_\text{N}^2$ crosses $0$, $x = 0.081 \pm 0.012$.
In (d), when $m_\text{str}^2$ crosses $0$, $x = 0.546 \pm 0.020$.
In (f), over $x \in [0.08, 0.55]$, the average value of $q_\text{EA}$ is $0.017 \pm 0.004$.
}
\label{fig:order}
\end{figure}

Throughout this work, we present a comprehensive ED study of the model with quantum spins. We take Heisenberg AFM coupling strengths obtained from quantum chemistry calculations~\cite{henrik2020e}:
\begin{equation}
(J_1,\ J_2,\ J'_1,\ J'_2,\ J''_1) = (7.4,\ 8.3,\ 0.7,\ 0.1,\ 0.3), \label{eq:mol-set}
\end{equation}
in the unit of meV.
Remarkably, as shown in \cref{fig:model}~(b) and \cref{fig:order}~(a)--(d), from the order parameters we identify a nonmagnetic ground state for Sr$_2$CuTe$_{1-x}$W$_x$O$_6$ in a wide range of $x \in [0.08, 0.55]$, matching well with the NMR observations~\cite{mustonen2018,mustonen2018t,watanabe2018} that there is no magnetic order over $19$~mK for $x \in [0.1, 0.6]$.
On top of this, the universal scaling exponent of the QFI density~\cite{zoller2016,scheie2021,laurell2021} extracted from the static spin structure factor vanishes near the phase boundaries, confirming the accuracy of our phase diagram.

The nature of the intermediate phase is distinct from normal candidates previously found in quantum magnets subject to quenched randomness, including the random-singlet state~\cite{kazuki2018,ren2023,liu2018,lee1982,fisher1994,baek2020}, the valence-bond glass~\cite{watanabe2018,tarzia2008,singh2010}, and the spin glass~\cite{kazuki2018,scheie2021,laurell2021}. Instead, its main features are more akin to short-range disordered SLL states in two dimensions, which have an extremely small spin-freezing parameter and a broad tail in the dynamical spin structure factor. Intriguingly, deep inside the nonmagnetic phase around $x = 0.3$ where these characteristics become the most predominant, the system displays zero averaged spin-spin correlation beyond nearest neighbors, and almost uniformly distributed static spin structure factor, leading to a uniform QFI density at $T = 0$.

The paper is structured as follows. In \cref{sec:model}, based on ED computations, we perform finite-size scaling of the order parameters and determine the phase diagram for the random-plaquette $J_1$-$J_2$ model. We characterize the emergent short-range disordered SLL state by spin-spin correlations, spin relaxations, as well as spin structure factors. In \cref{sec:qfi}, we focus on the QFI and reveal its universal scaling behaviors at quantum phase transitions. In \cref{sec:frus}, we advance the understanding of the phase formation in Sr$_2$CuTe$_{1-x}$W$_x$O$_6$ qualitatively through the randomness-induced frustration index.

\section{Properties of the model}
\label{sec:model}

Before proceeding to study physical observables, we first briefly introduce our reweighting method to compute the observables, which enables us to obtain the disorder-averaged ground-state expectation values with an arbitrary mixing ratio $x \in [0, 1]$. For each system size $N = 8, 12, 16, 20, 24, 28, 32$, we construct a lattice whose geometry allows both N{\'e}el and stripe orders under periodic boundary conditions (PBC). Then for each $k = 0, 1, \ldots, N$, we randomly generate $M_k$ Hamiltonian instances, each with exactly $k$ $J_2$-dominated plaquettes. As $x$ is the probability for a plaquette to be $J_2$-dominated in the disorder average, any physical observable $O$ is averaged under a binomial distribution, denoted by $[\,\cdots]_J$:
\begin{gather}
\left[\!\ev*{O} \right]_J(N, x) = \sum_k B(k, N, x) \bar{O}_k(N), \label{eq:bd} \\
\bar{O}_k(N) = \frac{1}{M_k} \sum_i \ev*{O}_i,
\end{gather}
where the binomial distribution takes the form $B(k, N, x) = \binom{N}{k} x^k (1 - x)^{N - k}$, and the binomial coefficient reads $\binom{N}{k} = N! / (k!\,(N - k)!)$. In $\bar{O}_k(N)$, the expectation $\ev*{O}_i$ acts on the ground state of the random instance $i$. Without reweighting, $\bar{O}_k(N)$ alone is an estimator of the uniform average over all random instances with a fixed $k$, under which circumstance $x$ can only take discrete values $k / N$ with a finite system size $N$. Our reweighting method makes it possible to vary $x$ continuously and more precisely determine the phase boundaries, as shown in the right column of \cref{fig:order}. More details on the ED computation, including choices of lattice geometries, reweighting, and estimation of uncertainties in the average and the extrapolation, can be found in \cref{app:ed}.

We comment on the symmetries of the ground state. It inherits the SU$(2)$ symmetry from Heisenberg exchanges, leading to zero magnetization $\ev*{S^z_j} = 0$ for each site $j$. The ground state also respects spin inversion symmetry and selects the even parity sector for all lattices we use, regardless of the value of $x$ and the configuration of random plaquettes. Translational symmetry and $D_4$ lattice symmetry, by contrast, are only present at $x = 0$ and $1$.

\subsection{Order parameters and phase diagram}
\label{sec:order}

Now, we determine the phase diagram of the random-plaquette $J_1$-$J_2$ model (see \cref{fig:model}) from four order parameters shown in \cref{fig:order}. While the N{\'e}el and the stripe order parameters set the boundaries of two magnetically ordered phases when $x$ is close to $0$ and $1$ respectively, the spin-freezing parameter and the valence-bond-glass parameter suggest that the nonmagnetic region $x \in [0.08, 0.55]$ cannot stabilize a random-singlet state nor a valence-bond glass.

First, let us take a closer look at two magnetically ordered phases. We choose the following definitions~\cite{kazuki2018} for the N{\'e}el and the stripe order parameters:
\begin{align}
m_\text{N}^2 &= \frac{8}{N (N + 4)} \left[ \sum_\alpha \sum_{i, j \in \alpha}\!\ev*{\bm{S}_i \cdot \bm{S}_j} \right]_J, \label{eq:op-n} \\
m_\text{str}^2 &= \frac{4}{N (N + 4)} \left[ \sum_\nu\,\sum_{\alpha_\nu} \sum_{i, j \in \alpha_\nu}\!\!\ev*{\bm{S}_i \cdot \bm{S}_j} \right]_J. \label{eq:op-as}
\end{align}
In $m_\text{N}^2$, $\alpha$ denotes each of the two sublattices formed by the N{\'e}el order, as shown in different colors in \cref{fig:model}~(b), and $i, j$ run over all sites in the sublattice $\alpha$.
In $m_\text{str}^2$, $\nu$ denotes each of the two directions of stripes, giving rise to two patterns of sublattices $\alpha_\nu$. \Cref{fig:model}~(b) depicts the horizontal stripes, and vertical otherwise.
The disorder average $[\,\cdots]_J$ is evaluated using our reweighting method in \cref{eq:bd}.
The normalization factors in \cref{eq:op-n,eq:op-as} are chosen according to relevant magnetic orderings of the classical Heisenberg antiferromagnet. For a perfect N{\'e}el order, the sum inside $[\,\cdots]_J$ in $m_\text{N}^2$ containing both $i = j$ and $i \ne j$ amounts to $2 \left( \frac{N}{2} \cdot \frac{3}{4} + \frac{N}{2} (\frac{N}{2} - 1) \cdot \frac{1}{4} \right) = N (N + 4) / 8$. The factor in $m_\text{str}^2$ is further multiplied by $2$ taking into account two types of stripes $\nu$. We then perform a finite-size scaling with $8 \le N \le 32$, as shown in \cref{fig:order}~(a)--(d). It turns out that both order parameters exhibit $N^{-1/2}$ scaling behaviors, consistent with the modified spin-wave analysis for Heisenberg antiferromagnet on finite square lattices~\cite{takahashi1987,song2011}:
\begin{equation}
\ev*{\bm{S}_i \cdot \bm{S}_j} = (-1)^{\alpha(i, j)} \left( c_0 + \frac{c_1}{|\bm{r}_i - \bm{r}_j|} \right)^2, \quad i \ne j, \label{eq:sw}
\end{equation}
where both $c_0$ and $c_1$ are constant given a finite $N$. The sign $(-1)^{\alpha(i, j)} = 1$ if the sites $i$ and $j$ belong to the same sublattice $\alpha$, and $-1$ otherwise.
\Cref{eq:sw} leads directly to the scaling relation:
\begin{equation}
m_\text{N/str}^2 = \frac{a}{\sqrt{N}} + b, \quad b \ge 0,
\end{equation}
where higher order corrections in $O(1 / N)$ are neglected.
In the thermodynamic limit $N \to \infty$, if the long-range magnetic order exists, $m_\text{N}^2$ and $m_\text{str}^2$ become positive constants.
From the non-negative extrapolation values of $m_\text{N}^2$ and $m_\text{str}^2$ in \cref{fig:order}~(b) and (d), we identify that in the ground state, the N{\'e}el phase survives from the randomness when $x \in [0, 0.08]$ and the stripe phase survives when $x \in [0.55, 1]$. The two quantum phase transition points crossing zero are resolved as $x_{c, 1} = 0.081 \pm 0.012$, $x_{c, 2} = 0.546 \pm 0.020$.
The extremely narrow region of the N{\'e}el order compared with the stripe order can be explained by reviewing the microscopic structure of the mixed compound in \cref{fig:model}~(a). When two different kinds of plaquette encounter, one strong $J_1$ bond is replaced by a much weaker $J''_1$ bond on the shared edge, while the strong $J_2$ bonds in the $J_2$-dominated plaquette remain unaffected.

The consistency of our nonmagnetic phase in absence of long-range order for $x \in [0.08, 0.55]$ with the experimental results $x \in [0.1, 0.6]$~\cite{mustonen2018,mustonen2018t,watanabe2018} is crucial. It validates the methodology of the current work, including the Hamiltonian we start with in \cref{eq:mol-set}, our ED computations, as well as our reweighting method in \cref{eq:bd} and extrapolation method. Next, we proceed to explore the nature of the nonmagnetic phase by testing two potential candidates: the random-singlet state and the valence-bond glass.

The random-singlet state can be characterized by the the spin-freezing parameter~\cite{kazuki2018,ren2023}, also known as the Edwards--Anderson (EA) order parameter~\cite{edwards1975,binder1986}, defined by
\begin{equation}
q_\text{EA}^2 = \frac{1}{N^2} \left[ \sum_{i, j} \ev*{\bm{S}_i \cdot \bm{S}_j}^2 \right]_J, \label{eq:sf}
\end{equation}
where $\sum_i$ sums over a total number of $N$ sites.
$q_\text{EA}$ detects any type of static spin order, including long-range magnetic orders and the spin-glass order.
In \cref{fig:order}~(e)--(f), $q_\text{EA}$ scales with $N^{-1/2}$ and extrapolates to positive values as $N \to \infty$ for all $x$, with the nonmagentic region $x \in [0.08, 0.55]$ locating a plateau around its minimum. The inset in (f) shows that the error bars of the extrapolation in the nonmagnetic phase do not extend to zero, and the average value of the plateau stays at $0.017 \pm 0.004$. Our finding agrees with the linear spin-wave analysis~\cite{henrik2022} of the same model in \cref{eq:mol-set} that the degree of frustration arising from weak Heisenberg interactions $J'_1, J'_2, J''_1$ is not strong enough to stabilize an all-to-all disordered random-singlet state in Sr$_2$CuTe$_{1-x}$W$_x$O$_6$.

The valence-bond glass (VBG), on the other hand, can be characterized by an order parameter that replaces the spin operator in $q_\text{EA}$ with the dimer operator~\cite{wu2019}:
\begin{gather}
q_\text{VBG}^2 = \frac{1}{4 N^2} \left[ \sum_{i, j} \sum_{p, q} \ev*{B_i^p B_j^q}^2 \right]_J, \\
B_i^p = \bm{S}_i \cdot \bm{S}_{i + p} - \ev*{\bm{S}_i \cdot \bm{S}_{i + p}},
\end{gather}
where $B_i^p$ encodes the dimer-dimer correlation~\cite{nomura2021} on a nearest-neighbor bond along the unit vector $p$, and $p$ can be either in $x$ or $y$ direction, i.e., $\sum_i \sum_p$ sums over a total number of $2 N$ nearest-neighbor bonds. As shown in \cref{fig:order}~(g)--(h), $q_\text{VBG}$ also exhibits $N^{-1/2}$ scaling behavior, but extrapolates to negative values as $N \to \infty$ for all $x$. We have also tested another dimer operator that sums over both nearest and next-nearest bonds and obtained the same behavior. Therefore, the putative valence-bond-glass state from the specific heat measurement at low temperatures~\cite{watanabe2018}, is not observed by our ED results of the random-plaquette model at $T = 0$.

\subsection{Short-range spin-spin correlations and non-glassy dynamics in the nonmagnetic phase}
\label{sec:sg}

\begin{figure}[t]
\centering
\includegraphics[width=\linewidth]{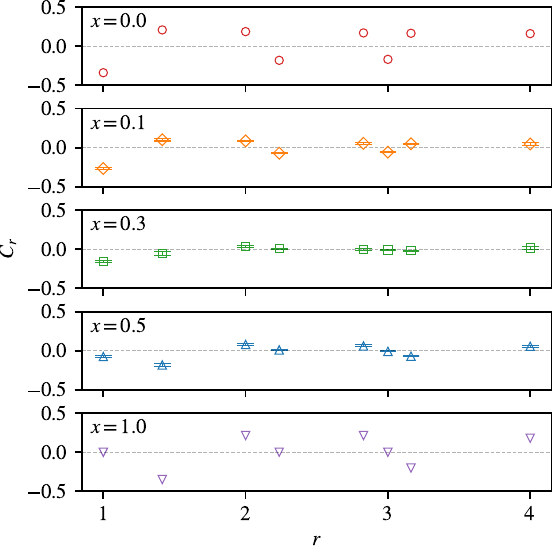}
\caption{
Averaged spin-spin correlation $C_r$ of the system with $N = 32$ spins. The length of the nearest-neighbor bond is set to $1$.
The error bars include standard errors over random instances and over all spin pairs $(i, j)$ sharing the same $r$ in an instance. At $x = 0$ and $1$, the zero error bar is caused by the absence of randomness and the restored $D_4$ lattice symmetry. Deep inside the nonmagnetic phase around $x = 0.3$, the spin-spin correlation vanishes beyond nearest neighbors.
}
\label{fig:corr_r}
\end{figure}

\begin{figure}[t]
\centering
\includegraphics[width=\linewidth]{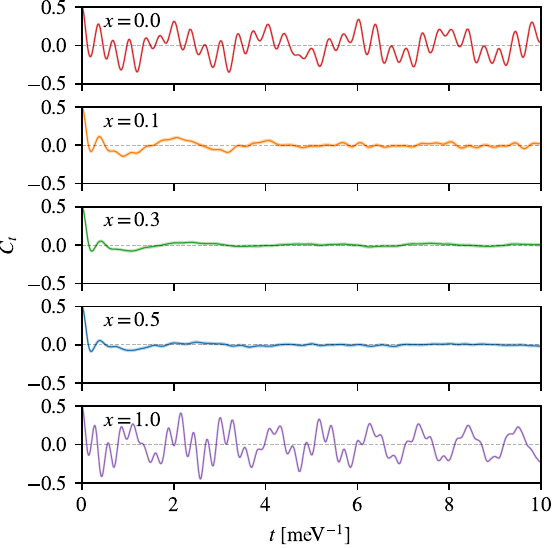}
\caption{
Temporal auto-correlation $C_t$ of a spin in the system with $N = 32$ spins. The shaded regions are standard errors over random instances. The dynamics is ergodic as the auto-correlation relaxes to zero quickly in the nonmagnetic phase $x \in [0.08, 0.55]$.
}
\label{fig:corr_t}
\end{figure}

A small but finite $q_\text{EA}$ when $N \to \infty$ could indicate a weak spin-glass order. However, as we will show in this section through spin-spin correlations over space and time domains, see \cref{fig:corr_r,fig:corr_t}, the spin glass is also not a candidate for the nonmagnetic phase.

First, let us examine the spin-spin correlation in real space,
\begin{equation}
C_r = \left[ \ev*{\bm{S}_i \cdot \bm{S}_j} \right]_{J, \bm{r}}, \quad r = |\bm{r}_i - \bm{r}_j|.
\end{equation}
Besides the disorder average $[\,\cdots]_J$ over all random instances in \cref{eq:bd}, we also take the average over all spin pairs $(i, j)$ sharing the same distance $r$ in each random instance~\cite{saha2023}.
\Cref{fig:corr_r} shows that while the N{\'e}el (stripe) order is indicated by the dominant negative correlations at nearest-neighbor $r = 1$ (next-nearest-neighbor $r = \sqrt{2}$), the long-range spin-spin correlations decrease quickly when $x$ enters the nonmagnetic phase. In particular, deep inside this phase around $x = 0.3$, we observe zero spin-spin correlation beyond nearest neighbors. The remaining AFM spin-spin correlations on nearest neighbors contribute to the non-zero extrapolation of spin-freezing parameter $q_\text{EA}$ in \cref{fig:order}~(e)--(f), and it is incompatible with the random domain wall picture of the spin-glass state in the random-bond $J_1$-$J_2$ model. According to Ref.~\cite{kazuki2018}, the latter would host a considerable amount of positive ferromagnetic (FM) correlations on nearest-neighbor spins across the domain walls between two distinct stripe-ordered states. In our model, the randomness-induced frustration plays a more important role, of which we leave the discussion to \cref{sec:frus}.

Next, we study the temporal auto-correlation of one spin in the ground state, which can probe the ergodicity breaking in finite-size glassy systems~\cite{carleo2012}:
\begin{equation}
C_t = \left[ \ev*{\{S^{z \dagger}(t), S^z(0)\}} \right]_J,
\end{equation}
where $S^{z \dagger}(t) = e^{i H t} S^z(0) e^{-i H t}$, and $C_t \in \mathbb{R}$ since $S^{z \dagger}(t) = S^z(t)$. It is noted that the choice of the spin $S^z$ in the system is arbitrary after averaging over random instances, and the $z$ orientation is also arbitrary as the model is $\text{SU}(2)$ symmetric. In \cref{fig:corr_t}, the two magnetically ordered phases at $x = 0$ and $1$ exhibit oscillating patterns that do not decay at large $t$. By contrast, in the nonmagnetic phase $x \in [0.08, 0.55]$, the auto-correlation relaxes to zero quickly. We do not observe the non-ergodic behavior in $C_t$ that is characteristic of a spin glass, which would entail a long-lived plateau at large $t$.

\subsection{Static and dynamical spin structure factors}
\label{sec:ssf}

\begin{figure}[t]
\centering
\includegraphics[width=\linewidth]{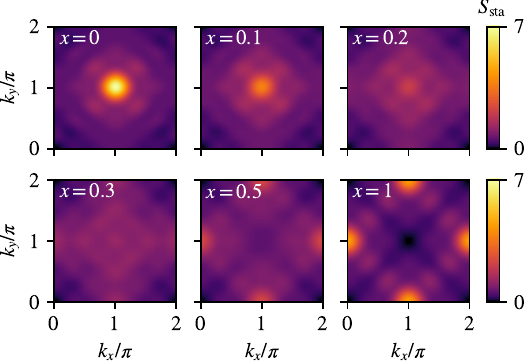}
\caption{
Static spin structure factor $S_\text{sta}$ of the system with $N = 32$ spins.
From \cref{eq:fq-sta}, it also displays the QFI density through the relation $f_\text{Q}(\bm{k}) = \frac{4}{3} S_\text{sta}(\bm{k})$.
}
\label{fig:static}
\end{figure}

After invalidating three candidates, namely random-singlet state, valence-bond glass, and spin glass, we proceed to unveil nontrivial SLL signatures of the nonmagnetic phase through spin structure factors.

The static spin structure factor defined by
\begin{equation}
S_\text{sta}(\bm{k}) = \frac{1}{N} \left[ \sum_{i, j} \ev*{\bm{S}_i \cdot \bm{S}_j} \cos\left( \bm{k} \cdot (\bm{r}_i - \bm{r}_j) \right) \right]_J,
\end{equation}
is resolved in the ground state via ED in \cref{fig:static}, in close resemblance to the experimental data probed by diffuse polarized neutron diffraction at $T = 1.5$~K~\cite{henrik2022}.
As $x$ increases from $0$ (N{\'e}el order) to $1$ (stripe order), the peak of $S_\text{sta}$ moves from $(\pi, \pi)$ to $(\pi, 0)$. Meanwhile, \cref{fig:static} shows a more detailed feature that the peak becomes less sharp once $x$ leaves the ordered phases and approaches the phase boundaries at $x = 0.1$ and $0.5$, which also agrees with the experimental data. Interestingly, around $x = 0.3$, $S_\text{sta}$ becomes almost uniform in a large portion of the Brillouin zone (BZ) without any visible peak. It indicates a strongly disordered state, absent of any long-range magnetic order.

\begin{figure}[t]
\centering
\includegraphics[width=\linewidth]{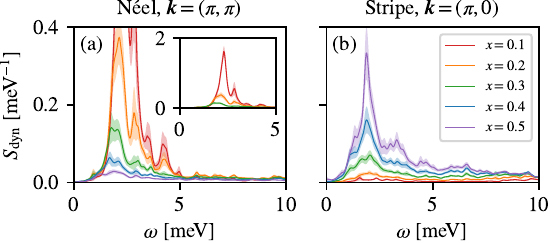}
\caption{
Dynamical spin structure factor $S_\text{dyn}$ of the system with $N = 32$ spins, exhibiting broad peaks at $\bm{k} = (\pi, \pi)$ and $(\pi, 0)$ in the nonmagnetic phase $x \in [0.08, 0.55]$. The shaded regions are standard errors over random instances.
}
\label{fig:ds}
\end{figure}

Furthermore, as shown from the dynamical spin structure factor in \cref{fig:ds}, the coexistence of a peak centered at a small frequency $\omega$ and a broad tail at large $\omega$ also supports a SLL state~\cite{kazuki2018,wu2019} in the nonmagnetic region around $x = 0.3$. In our calculation, we adopt the following definition for the dynamical spin structure factor:
\begin{align}
S_\text{dyn}(\bm{k}, \omega)
&= \frac{1}{2 \pi} \int_{-\infty}^\infty \left[ \ev*{S^{z \dagger}_{\bm{k}}(t) S^z_{\bm{k}}(0)} \right]_J\!e^{i \omega t} \dd t \label{eq:sd} \\
&= -\lim_{\eta \to 0} \left[ \frac{1}{\pi} \Im \ev*{S^{z \dagger}_{\bm{k}} \frac{1}{\omega + E_0 - H + i \eta} S^z_{\bm{k}}} \right]_J, \notag
\end{align}
with $E_0$ the ground state energy.
The time-domain integral is computed using continued fraction expansion~\cite{gagliano1987} and smoothed with a sufficiently small damping factor $\eta = 0.1$~meV~\cite{kazuki2018, wu2019}. The Fourier transform of a spin operator is taken to be
\begin{equation}
S^z_{\bm{k}} = \frac{1}{\sqrt{N}} \sum_{i = 1}^N S_i^z e^{-i \bm{k} \cdot \bm{r}_i}.
\end{equation}
\Cref{fig:ds} shows that when the mixing ratio $x$ increases in the nonmagnetic region, the peak of $S_\text{dyn}$ moves in the BZ from the center $(\pi, \pi)$ to the edge $(\pi, 0)$. In the frequency domain, the peak moves towards smaller $\omega$ as the randomness $x$ increases at $\bm{k} = (\pi, \pi)$, while at $(\pi, 0)$ it stays rather fixed near $\omega \approx 2$~meV. Most importantly, taking into account the flatness of the static spin structure factor in the nonmagnetic phase in \cref{fig:static}, the broad peak of $S_\text{dyn}$ extends naturally to other wave vectors in the BZ. This behavior can be used to distinguish from magnon-like excitations which introduce peaks only at high-symmetry points of the BZ~\cite{wu2019}.
We have also tested finite-size effects and verified that for a fixed $x \in [0.08, 0.55]$ at both wave vectors, the peak reaches lower frequency as the system size increases.

\section{Quantum Fisher information in 2D random quantum magnets}
\label{sec:qfi}

In this section, we introduce the general relation between the QFI and the spin structure factors, and demonstrate how to use the QFI density as a witness of multipartite entanglement in the random-plaquette $J_1$-$J_2$ model.

\subsection{Multipartite entanglement witness}

Let us start from a generic equilibrium mixed state at the temperature $T$. Its density matrix has the form $\rho = \sum_\lambda p_\lambda \ket{\lambda}\bra{\lambda}$ with the probability $p_\lambda = \frac{1}{Z} e^{-E_\lambda / T}$ and the partition function $Z = \sum_\lambda e^{-E_\lambda / T}$, where we set $\hbar = k_B = 1$. Given a generator $O = \sum_{i = 1}^N O_i$ consisting of local operators $O_i$ that are normalized with unit spectrum width, the QFI is defined by
\begin{equation}
f_\text{Q}(T) = 2 \sum_{\lambda, \lambda'} \frac{(p_\lambda - p_{\lambda'})^2}{p_\lambda + p_{\lambda'}} \left|\!\mel{\lambda}{O}{\lambda'} \right|^2,
\label{eq:fq_sum}
\end{equation}
with $p_\lambda + p_{\lambda'} > 0$.
Physically, through the quantum Cram{\'e}r--Rao bound~\cite{braunstein1994}, $f_\text{Q}$ quantifies the maximal precision of a parameter $\theta$ one can reach after
$M$ independent measurements: $\Var \theta \ge 1 / (M f_\text{Q})$.
It is a lower bound for the variance of parameter $\theta$ associated with the generator $O$. Under each measurement, the system undergoes a unitary transformation: $\rho' = e^{-i \theta O} \rho e^{i \theta O}$. The QFI has been successfully measured in a single spin system, or a solid-state qubit~\cite{nathan2022}.

Going to multispin systems of higher complexity, an exact relation can be established~\cite{zoller2016,scheie2021,laurell2021} between the QFI density $f_\text{Q} = f_\text{Q} / N$ and the dynamical spin structure factor $S_\text{dyn}$:
\begin{equation}
f_\text{Q}(\bm{k}, T) = 4 \int_0^\infty\!\!\tanh(\frac{\omega}{2 T}) \left( 1 - e^{-\frac{\omega}{T}} \right) S_\text{dyn}(\bm{k}, \omega, T)\,\dd \omega,
\label{eq:fq}
\end{equation}
with the generator chosen to be $O(\bm{k}) = \sqrt{N} S^{z \dagger}_{\bm{k}} = \sum_{i = 1}^N S_i^z e^{i \bm{k} \cdot \bm{r}_i}$. Here, $S_\text{dyn}$ is obtained from \cref{eq:sd} by applying $\ev*{S^{z \dagger}_{\bm{k}}(t) S^z_{\bm{k}}(0)} = \Tr[\rho\,S^{z \dagger}_{\bm{k}}(t) S^z_{\bm{k}}(0)]$, which acts on the thermal state when $T > 0$ and on the ground state at $T = 0$. \Cref{eq:fq} expresses the entanglement in an explicit form of quantum fluctuations. A detailed derivation is provided in \cref{app:qfi_rev}.

Notably, a quantum state is $(m + 1)$-partite entangled when the QFI density satisfies the inequality~\cite{zoller2016,laurell2021,hyllus2012,toth2012}: $f_\text{Q} > m (h_\text{max} - h_\text{min})^2$, where $m$ is an integer and $h_\text{max}$ ($h_\text{min}$) is the largest (smallest) eigenvalue of the local generator $O_i$. In our case, $\ev*{O^\dagger_i O_i} = \ev*{(S_i^z)^2} = 1/4$ and $h_\text{max} = -h_\text{min} = 1/2$. Therefore, $(m + 1)$-partite entanglement is witnessed in a multispin system if
\begin{equation}
f_\text{Q} > m. \label{eq:lb}
\end{equation}
At the same time, \cref{eq:lb} sets the criterion according to which the quantum Cram{\'e}r--Rao bound falls below the classical limits~\cite{pezze2014}.

\begin{figure}[t]
\centering
\includegraphics[width=\linewidth]{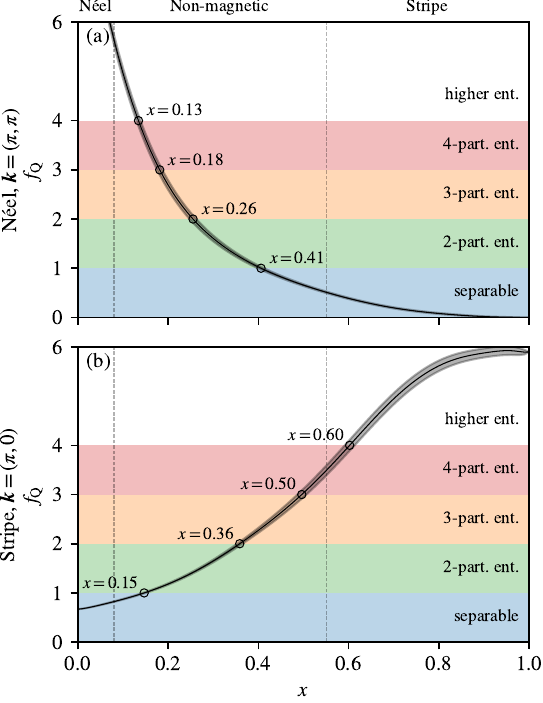}
\caption{
QFI density $f_\text{Q}$ and multipartite entanglement of the system at $T = 0$ with $N = 32$ spins. The vertical dashed lines indicate the nonmagnetic phase $x \in [0.08, 0.55]$. The shaded regions are standard errors over random instances.
}
\label{fig:qfi-gs}
\end{figure}

Next, we focus on the features of the QFI in the ground state of the random-plaquette model. By taking the limit $T \to 0$ in \cref{eq:fq}, we can verify that
\begin{equation}
f_\text{Q}(\bm{k}) = 4 \int_0^\infty S_\text{dyn}(\bm{k}, \omega)\,\dd \omega = \frac{4}{3} S_\text{sta}(\bm{k}). \label{eq:fq-sta}
\end{equation}
The extra factor $1/3$ comes from one of the three $\text{SU}(2)$-symmetric components $\bm{S}_i \cdot \bm{S}_j = \sum_{\alpha = x, y, z} S_i^\alpha S_j^\alpha$.
Equivalently, we can also evaluate the QFI of a pure ground state $\ket{\psi}$. The density matrix becomes $\rho = \ket{\psi}\bra{\psi}$, and the case $p_\psi = 1$, $p_{\psi'} = 0$ contributes twice to $f_\text{Q}$ in \cref{eq:fq_sum}. Applying the identity $\ket{\psi'}\bra{\psi'} = \mathds{1} - \ket{\psi}\bra{\psi}$, we obtain the same relation
\begin{equation}
f_\text{Q}(\bm{k}) = 4 \left[ \ev*{S^z_{\bm{k}} S^{z \dagger}_{\bm{k}}} - |\ev*{S^z_{\bm{k}}}|^2 \right]_J = \frac{4}{3} S_\text{sta}(\bm{k}).
\end{equation}
It is noted that $\ev*{S^z_{\bm{k}}} = 0$ respecting the $\text{SU}(2)$ symmetry.

In \cref{fig:qfi-gs}, we quantify multipartite entanglement witnessed by $f_\text{Q}$ at the two ordering wave vectors using \cref{eq:lb,eq:fq-sta}. The nonmagnetic phase is short-range entangled, featuring $2$- to $4$-partite entanglement on a finite-size lattice of $N = 32$ spins. By contrast, the N{\'e}el and the stripe phases possess much higher entanglement at their respective ordering wave vectors. This long-range entanglement is also reflected in the sharp peaks of $S_\text{sta}(\bm{k})$ in \cref{fig:static}.
When the mixing ratio $x$ increases from $0$ to $1$, the peak of $f_\text{Q}$ moves from $\bm{k} = (\pi, \pi)$ to $(\pi, 0)$ following the same result as $S_\text{sta}$. Around $x = 0.3$ deep inside the nonmagnetic phase, $f_\text{Q}$ also becomes uniform and the system displays bipartite entanglement over the BZ.

\subsection{Universal scaling at quantum critical points}

\begin{figure}[t]
\centering
\includegraphics[width=\linewidth]{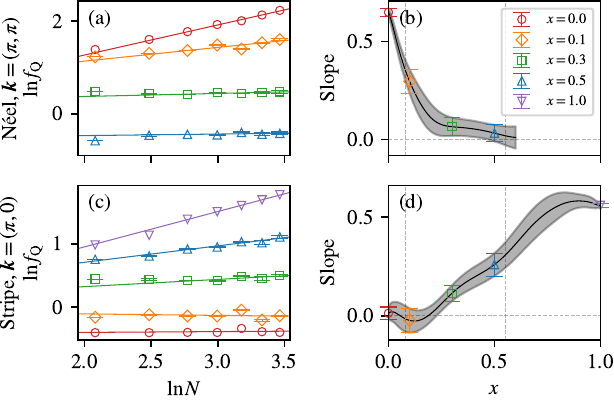}
\caption{
Scaling of QFI density $f_\text{Q}$ at $T = 0$.
(a) (c) Linear regression of $\ln f_\text{Q} = \nu_\text{Q} \ln N + c_0$ at specific values of $\bm{k}$ and $x$, where $\nu_\text{Q}$ is the scaling exponent and $c_0$ is a constant.
(b) (d) The slope $\nu_\text{Q}$ as a function of $x$. The shaded regions are standard errors of $\nu_\text{Q}$. The horizontal dashed line indicates $\nu_\text{Q} = 0$, and the vertical dashed lines indicate the nonmagnetic phase $x \in [0.08, 0.55]$.
In (a) (b), only the results of $x < 0.6$ are shown, where $f_\text{Q}$ is large enough to numerically estimate the slope.
In (b), when $x_\text{c} = 0.55$, $\nu_\text{Q} = 0.02 \pm 0.05$.
In (d), when $x_\text{c} = 0.08$, $\nu_\text{Q} = -0.02 \pm 0.06$.
}
\label{fig:qfi-scal}
\end{figure}

Another interesting feature observed from \cref{fig:qfi-gs} is that in the finite-size system with $N = 32$, if we choose an ordering wave vector and tunes $x$ towards the other ordered phase, there is always no multipartite entanglement to be witnessed at the phase boundary, as $f_\text{Q} < 1$ at $\bm{k} = (\pi, \pi)$, $x = 0.55$ and $\bm{k} = (\pi, 0)$, $x = 0.08$. In the following, we will demonstrate through the scaling analysis of $f_\text{Q}$ that in the thermodynamic limit, the critical points for $f_\text{Q} = 1$ before entering a separable (unentangled) region coincide with the phase boundaries of the random-plaquette model determined by the order parameters.

At $T = 0$, we start from a general scaling relation between the QFI density and the system size:
\begin{equation}
\ln f_\text{Q} (\bm{k}) = \nu_\text{Q} (\bm{k}) \ln N + c_0,
\end{equation}
where $\nu_\text{Q}$ denotes the scaling exponent and $c_0$ is a constant.
If $\nu_\text{Q} = 0$, then $f_\text{Q}$ converges to a finite value when we take the thermodynamic limit $N \to \infty$. The finite-size ED computations on our model lead to \cref{fig:qfi-scal}. At $\bm{k} = (\pi, \pi)$, $\nu_\text{Q}$ is positive when $x < 0.55$ and vanishes as $\nu_\text{Q} = 0.02 \pm 0.05$ around the critical point $x_\text{c} = 0.55$. Thus when $N \to \infty$, $f_\text{Q}$ converges to a finite value $f_\text{Q} \approx 1$ only when $x$ is around the critical point.
The same behaviour of $f_\text{Q}$ is observed at $\bm{k} = (\pi, 0)$, which exhibits a vanishing exponent $\nu_\text{Q} = -0.02 \pm 0.06$ around the critical point $x_\text{c} = 0.08$.

To conclude, we find that the QFI density is capable to detect the quantum phase transitions at $T = 0$ in the random-plaquette model. Given an ordering wave vector $\bm{k}$, the scaling exponent of $f_\text{Q}$ with respect to the system size vanishes across the transition point $x_\text{c}$:
\begin{equation}
\left. \nu_\text{Q}(\bm{k}) \right|_{x = x_\text{c}} \approx 0, \label{eq:vq}
\end{equation}
which signifies the other magnetically ordered phase. It is noted that the scaling exponent of $f_\text{Q}$ also vanishes at topological phase transitions, e.g. in the Kitaev superconducting chain~\cite{zoller2016}. Through \cref{eq:vq}, we extend the universality of this phenomenon to the order-disorder phase transitions in 2D random quantum magnets.

\section{Statistics of randomness-induced frustration}
\label{sec:frus}

\begin{figure}[t]
\centering
\includegraphics[width=\linewidth]{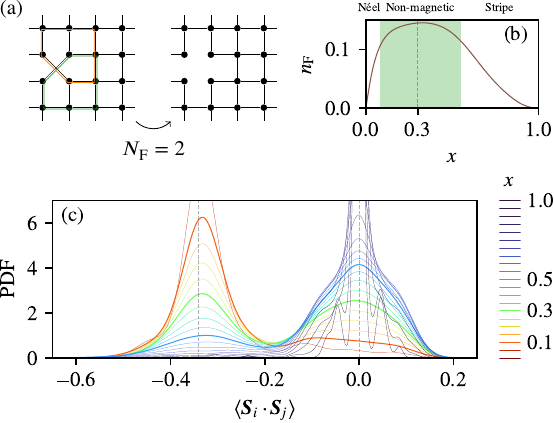}
\caption{
(a) An example of a graph containing only the large couplings $J_1$ and $J_2$ with $N = 16$ spins and one $J_2$-dominated plaquette. Two examples of frustrated cycles of spin couplings are highlighted in green and orange respectively. This graph becomes non-frustrated if at least two edges are removed, therefore it has frustration index $N_\text{F} = 2$.
(b) The normalized frustration index $n_\text{F} = N_\text{F} / N$ as a function of $x$, averaged over random instances with $N = 32$ spins. The green region indicates the nonmagnetic phase $x \in [0.08, 0.55]$. The maximum of $n_\text{F}$ is around $x = 0.3$, as indicated by the vertical dashed line.
(c) Distribution of nearest-neighbor spin correlations $\ev*{\bm{S}_i \cdot \bm{S}_j}$ with $N = 32$ spins. The results of $x = 0.1, 0.3, 0.5$ are highlighted with thick curves. When $x = 0$ and $1$, the distribution is fully concentrated at $\ev*{\bm{S}_i \cdot \bm{S}_j} = -0.340$ and $0$ respectively, as indicated by the vertical dashed lines. The probability density function (PDF) of the distribution is estimated from random instances using Silverman’s method~\cite{silverman1986density}.
}
\label{fig:fru}
\end{figure}

Extra insights can be gained on the role of the mixing ratio $x$ in Sr$_2$CuTe$_{1-x}$W$_x$O$_6$ by introducing the frustration index of a random instance. Illustrated in \cref{fig:fru}~(a), for each random instance of our model in \cref{eq:mol-set}, we consider a graph whose edges only contain the large couplings $J_1$ and $J_2$, and omit the small ones $J'_1$, $J'_2$, and $J''_1$. The frustration index $N_\text{F}$ is defined to be the minimal number of edges one needs to remove to make the graph geometrically non-frustrated (containing only even-length cycles of AFM couplings). In this simple example, we have $N_F = 2$. In generic cases, the method to compute it is equivalent to finding the ground state of a classical AFM Ising model on the graph, and to solving other well-studied combinatorial optimization problems such as max-cut~\cite{aref2020modeling}.

In the random-plaquette model, it is more convenient to consider the normalized frustration index $n_\text{F} = N_\text{F} / N$, which converges in the thermodynamic limit $N \to \infty$. Shown in \cref{fig:fru}~(b), when $n_\text{F}$ is small ($< 0.1$), the system exhibits the N{\'e}el or the stripe order. Each frustrated edge raises the ground state energy approximately by $2 J |S| (|S| + 1)$, where $J = J_1$ or $J_2$ depending on the position of the frustrated edge, and $|S| = 1/2$. Remarkably, when $x$ enters the nonmagnetic region $[0.08, 0.55]$, a high plateau of $n_\text{F}$ appears with a peak around $x = 0.3$, where it has so far been identified with most significant disordered SLL signatures. We can thus gain a phenomenological understanding on the origin of the nonmagnetic phase in Sr$_2$CuTe$_{1-x}$W$_x$O$_6$: The mixing ratio $x$ induces additional frustration, which plays a more important role in driving the system into a short-range disordered SLL phase than the originally present weak couplings $J'_1$, $J'_2$, and $J''_1$.

Moreover, in \cref{fig:fru}~(c) we show the distribution of nearest-neighbour spin correlations of the random-plaquette model at varying $x$ for a system with $N = 32$ spins. The lower and the upper bounds $\ev*{\bm{S}_i \cdot \bm{S}_j} \in [-3/4, 1/4]$ are given by pure spin singlet and triplet pairs. The two sharp peaks at $\ev*{\bm{S}_i \cdot \bm{S}_j} = -0.340$ and $0$ appear in the N{\'e}el and the stripe phases respectively. There is a smooth transfer of probability density between them as $x$ changes. Interestingly, in the nonmagnetic region $x \in [0.08, 0.55]$, the distribution spans much broader, and at $x = 0.3$ there appear two peaks of almost equal height.
Compared with the random-bond $J_1$-$J_2$ model~\cite{kazuki2018}, we do not observe any major peak of $\ev*{\bm{S}_i \cdot \bm{S}_j} \in [-0.2, -0.1]$ as in random-singlet states, and there is also no visible peak at $\ev*{\bm{S}_i \cdot \bm{S}_j} > 0$ as in spin-glass states. This distinguishes our nonmagnetic state from these two candidates.

\section{Discussion}
\label{sec:disc}

To summarize, in this work we perform extensive ED studies of the random-plaquette $J_1$-$J_2$ model with experimentally relevant parameters for the material Sr$_2$CuTe$_{1-x}$W$_x$O$_6$. We resolve the phase boundaries of a nonmagnetic state $x \in [0.08, 0.55]$ by the finite-size scaling of various order parameters. The intermediate phase is  absent of long-range N{\'e}el and stripe orders, valence-bond glass order, as well as glassy dynamics. Yet, it still exhibits short-range spin-spin correlations and a broad tail in $S_\text{dyn}$, which are characteristic of disordered SLL states. Deep inside the nonmagnetic phase around $x = 0.3$, the randomness-induced frustration index reaches the maximum, bringing a series of non-trivial features including almost uniform $S_\text{sta}$ (or equivalently, $f_\text{Q}$) over the BZ. The broad peaks in $S_\text{dyn}$ are also stretched away from the two ordering wave vectors $\bm{k} = (\pi, \pi)$ and $(\pi, 0)$, marking the difference from magnon-like excitations.

Although the bond frustration from the small couplings $J'_1, J'_2, J''_1$ is not strong enough to support an all-to-all disordered random-singlet state as $q_\text{EA} > 0$, the frustration index in the model increases with structured randomness, thus favoring a disordered SLL state. Finally, from the analysis of the QFI, the nonmagnetic phase is multipatite entangled, and the vanishing universal scaling exponent of the QFI density with respect to $N$ across phase transitions shows the robustness of our phase boundaries in the thermodynamic limit.

In future, one may generalize the calculation of the QFI to thermal equilibrium mixed states for $T > 0$, using \cref{eq:fq} and the Hams--de~Raedt method~\cite{hams2000}. A perfect scaling collapse at low temperatures~\cite{zoller2016} is expected from vanishing critical exponent at $T = 0$, and will extend the critical-$T$ regime of the phase diagram in \cref{fig:model}~(b). It is also possible to study the QFI far from equilibrium in our model using small-size ED clusters, e.g. $N = 10$ spins~\cite{baykusheva2023}.

Going beyond ED, one may also apply variational Monte Carlo (VMC) to explore 2D random quantum magnets. The recently developed neural quantum states~\cite{carleo2017,carleo2019}, especially the transformer architectures~\cite{viteritti2023,viteritti2023t}, can be expressive variational ansatzes to capture the rich complexity of the physical system. There have been successful attempts at applying neural quantum states to approximate the QSL state in the regular $J_1$-$J_2$ Heisenberg model on the square lattice~\cite{nomura2021,macdonald2023,markus2023,rende2023}, despite the nature of Dirac spin liquid is still under debate~\cite{senthil2024}.
Given the flexibility of neural networks, the access to larger system sizes is envisioned, which enables one to further resolve the dispersion of the dynamical spin structure factor and the spin excitation gaps, casting more light on the nature of the nonmagnetic state.

\section*{Acknowledgements}

We thank Federico Becca, Emil J.\ Bergholtz, Ellen Fogh, Yuchi He, Riccardo Rossi, and Luciano L.\ Viteritti for discussions. We especially thank Henrik M.\ R{\o}nnow for pointing our attention to the disordered model studied in this work.
Support from the Swiss National Science Foundation is acknowledged under Grant No.~200336.

\appendix

\setcounter{equation}{0}
\setcounter{figure}{0}
\setcounter{table}{0}

\renewcommand{\theequation}{S\arabic{equation}}
\renewcommand{\thefigure}{S\arabic{figure}}
\renewcommand{\thetable}{S\arabic{table}}

\section{Finite-size scaling on frustrated random magnets}
\label{app:ed}

\subsection{Choice of finite-size lattices}
\label{app:lattice}

We choose a series of lattices with $N \le 32$ to obtain enough data points from ED and extrapolate to $N \to \infty$, while maintaining the desired symmetries. Here we formalize the construction of such lattices.

\begin{figure*}[t]
\centering
\includegraphics[width=0.8\linewidth]{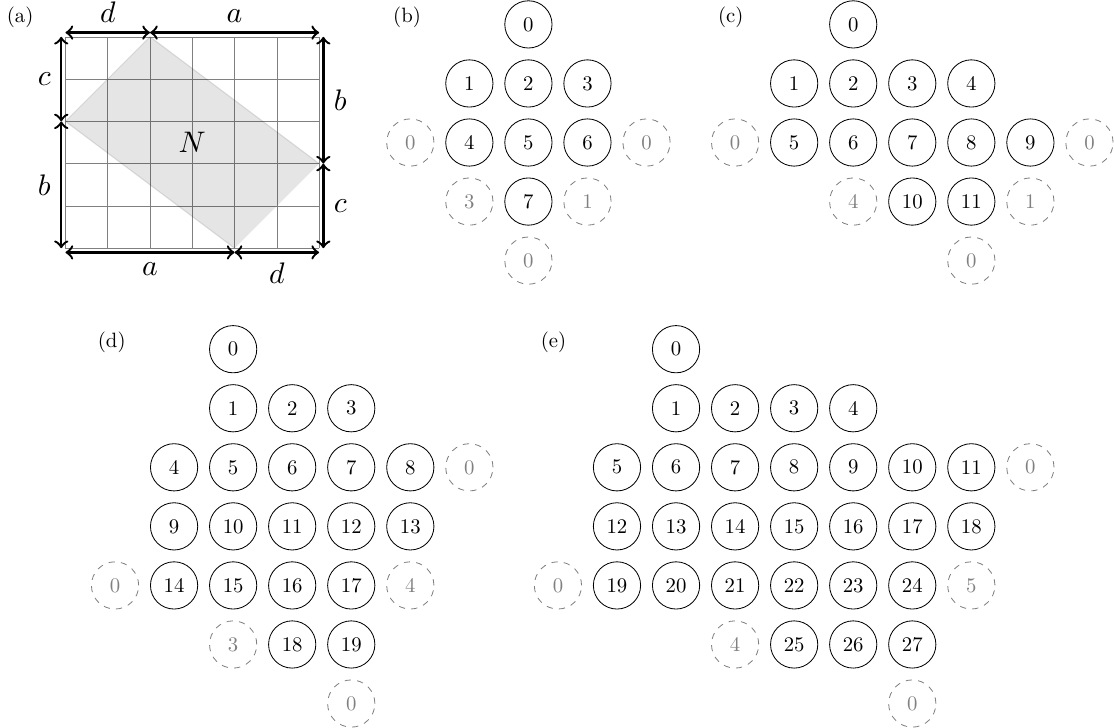}
\caption{
(a) Shape of a parallelogram unit cell.
(b)--(e) The lattices used with $N = 8, 12, 20, 28$. The sites are numbered for the ease of programmatic use. The dashed circles with the same number are taken as the same site in adjacent unit cells under PBC.
}
\label{fig:lattice}
\end{figure*}

The infinite square lattice can be covered by repeating parallelogram unit cells, whose shape is determined by the parameters $a$, $b$, $c$, $d$, and the number of lattice sites in each unit cell is $N = a c + b d$, as shown in \cref{fig:lattice}~(a). We denote such a parallelogram unit cell with PBC by the \emph{parallelogram lattice}.

We want the parallelogram lattices to be as close to square lattices as possible. The quantities $\delta_1 = |a^2 + b^2 - c^2 - d^2|$ and $\delta_2 = |a d - b c|$ are invariant under translations and reflections, where $\delta_1 = 0$ only if the four edges of the parallelogram have the same length, and $\delta_2 = 0$ only if the four angles are right. We heuristically define the \emph{non-squareness} $\delta = \frac{1}{4} \delta_1 + \delta_2$.

Not all the lattices have the topology to form the N{\'e}el and the stripe orderings. We denote a lattice is \emph{N{\'e}el-compatible} if it is bipartite, and \emph{stripe-compatible} if it has exactly two components (connected subgraphs that are mutually disconnected) with the same size and each component is bipartite. In corollary, a parallelogram lattice is N{\'e}el-compatible only if $a + b$ and $c + d$ are both even, and it is stripe-compatible only if $a$, $b$, $c$, $d$ are all even.

For each system size $N$, we choose the lattice that is both N{\'e}el- and stripe-compatible, and has minimal $\delta$. The chosen lattices are listed in \cref{tab:lattice}, and the non-trivial shapes are shown in \cref{fig:lattice}.

\begin{table}[b]
\centering
\begin{tabular}{r@{\hskip 1em}c@{\hskip 1em\!}r}
\toprule
$N$ & $(a, c, b, d)$ & $\delta$ \\
\midrule
8 & (2, 2, 2, 2) & 0 \\
12 & (4, 2, 2, 2) & 7 \\
16 & (4, 4, 0, 0) & 0 \\
20 & (4, 4, 2, 2) & 0 \\
24 & (6, 4, 0, 0) & 5 \\
28 & (6, 4, 2, 2) & 9 \\
32 & (4, 4, 4, 4) & 0 \\
\bottomrule
\end{tabular}
\caption{The parallelogram lattices used in this paper.}
\label{tab:lattice}
\end{table}

\subsection{Details of ED computation}

The ED is implemented using the software package \texttt{lattice-symmetries}~\cite{westerhout2021latticesymmetries}, which is one of the most performant among available software of the kind to our knowledge. We apply zero total magnetization ($\text{SU}(2)$ symmetry) and spin inversion symmetry to reduce the Hilbert space, and we have verified that all random instances have ground states in the even parity sector under spin inversion. Note that no translational or reflection symmetry can be applied because of the random plaquettes.

\subsection{Averaged observables over random instances}
\label{app:reweight}

We use stratified sampling to generate random instances of the Hamiltonian with various numbers of $J_2$-dominated plaquettes, then reweight them according to the binomial distribution when computing the averaged observables given the probability $x$. Therefore, we can reuse the observable values of the instances for all values of $x$, without recomputing the ED and the expectations.

For each $k = 0, 1, \ldots, N$, we generate $M_k = \min(M_{k \text{max}}, \binom{N}{k})$ instances randomly without replacement, each with exactly $k$ $J_2$-dominated plaquettes. Here $M_{k \text{max}}$ is the maximal number of instances for each $k$ limited by the computation budget, and the actual number $M_k$ is also limited by the number of combinations $\binom{N}{k}$. We take $M_{k \text{max}} = 100$ when $N \le 20$, $M_{k \text{max}} = 30$ when $N = 24, 28$, and $M_{k \text{max}} = 10$ when $N = 32$.

When computing an averaged observable $[O]_J = [\!\ev*{O}]_J$, we first take a uniform average over instances with the same $k$, then average over the binomial distribution of $k$:
\begin{gather}
[O]_J(N, x) = \sum_k B(k, N, x) \bar{O}_k(N), \\
B(k, N, x) = \binom{N}{k} x^k (1 - x)^{N - k}, \\
\bar{O}_k(N) = \frac{1}{M_k} \sum_i O_i,
\end{gather}
where $i$ runs over all $M_k$ instances whose number of $J_2$-dominated plaquettes equals $k$, and $O_i = \ev*{O}_i$ is the observable value of this instance. As $O_i$ is independent of $x$, we can reuse it for all values of $x$. For each $x$, we compute $[O]_J(N, x)$ with finite values of $N$, then extrapolate to $[O]_J(N \to \infty, x)$, as in \cref{fig:order}.

The standard error of this weighted average is given by
\begin{gather}
O_\text{err}^2(N, x) = \frac{1}{M_\text{eff}(N, x)} \sum_k B(k, N, x) O_{k \text{err}}^2(N), \\
M_\text{eff}(N, x) = \frac{1}{\sum_k B(k, N, x)^2}, \\
O_{k \text{err}}^2(N) = \left( 1 - \frac{M_k}{\binom{N}{k}} \right) \frac{1}{M_k^2} \sum_i (O_i - \bar{O}_k)^2,
\end{gather}
where $M_\text{eff}$ is the effective sample size over the binomial distribution, and $M_k / \binom{N}{k}$ is the finite population correction. The error bars $O_\text{err}$ are shown in \cref{fig:order} when $N < \infty$ and $0 < x < 1$.

\subsection{Uncertainty of extrapolation}

We use linear regression to extrapolate observables to $N \to \infty$, and propagate uncertainties of both the averages over random instances and the regression.

Consider a linear regression with $P$ data points $\bm{x} = (x_1, \ldots, x_P)^\top$, $\bm{y} = (y_1, \ldots, y_P)^\top$. The standard errors in $\bm{y}$ are $\bm{y}_\text{err} = (y_{\text{err} 1}, \ldots, y_{\text{err} P})^\top$, and there is no error in $\bm{x}$. We have
\begin{equation}
\bm{y} = \bm{X} \bm{\beta} + \bm{\epsilon},
\end{equation}
where $\bm{X} = (\bm{1}, \bm{x})$ is a $P \times 2$ matrix including the constant term, $\bm{\beta} = (\beta_0, \beta_1)^\top$ contains the intercept and the slope, and $\bm{\epsilon}$ is the errors of regression. Note that $\bm{\epsilon}$ is heteroscedastic and we cannot assume that all components of $\bm{\epsilon}$ are from distributions with the same variance, because the deviation of $\bm{y}$ from the linear relation is caused by the finite size effect, whose strength changes with the system size.

The parameters are estimated by least squares:
\begin{equation}
\bm{\beta} = \bar{\bm{X}} \bm{y}, \quad
\bar{\bm{X}} = (\bm{X}^\top \bm{X})^{-1} \bm{X}^\top,
\end{equation}
where $\bar{\bm{X}}$ is the pseudoinverse of $\bm{X}$. The variance matrix of parameters from the errors in $\bm{y}$ is
\begin{equation}
\Var^{(1)} \bm{\beta} = \bar{\bm{X}} (\diag \bm{y}_\text{err})^2 \bar{\bm{X}}^\top,
\end{equation}
and another contribution from the errors of regression is
\begin{gather}
\Var^{(2)} \bm{\beta} = \frac{P}{P - 2} \bar{\bm{X}} (\diag \bm{\epsilon})^2 \bar{\bm{X}}^\top, \\
\bm{\epsilon} = \bm{y} - \bm{X} \bm{\beta}.
\end{gather}
The total variance $\Var \bm{\beta} = \Var^{(1)} \bm{\beta} + \Var^{(2)} \bm{\beta}$, and the standard error of the intercept $\beta_{0 \text{err}}$ is the square root of its first component. The error bars $\beta_{0 \text{err}}$ are shown in \cref{fig:order} when $N \to \infty$.

\section{QFI density from dynamical spin structure factor}
\label{app:qfi_rev}

Here, we briefly review the approach~\cite{zoller2016,scheie2021,laurell2021} of obtaining the QFI density $f_\text{Q}$ through the dynamical spin structure in \cref{eq:fq}. It assumes thermal ensembles at the equilibrium and adapts the fluctuation-dissipation theorem for non-symmetrized $S_\text{dyn}(\bm{k}, \omega, T)$~\cite{lovesey1984}.

Let us take a generator $O = \sqrt{N} S^{z \dagger}_{\bm{k}} = \sum_{i = 1}^N S_i^z e^{i \bm{k} \cdot \bm{r}_i}$. Considering the definition of QFI density $f_\text{Q} = f_\text{Q} / N$ and the form of QFI in \cref{eq:fq_sum}, we have
\begin{equation}
f_\text{Q}(\bm{k}, T) = 2 \sum_{\lambda, \lambda'} \frac{(p_\lambda - p_{\lambda'})^2}{p_\lambda + p_{\lambda'}} \left|\!\mel{\lambda}{S^{z \dagger}_{\bm{k}}}{\lambda'} \right|^2.
\label{eq:f0}
\end{equation}
It is convenient to introduce the dynamical susceptibility
\begin{equation}
\chi(\bm{k}, \omega, T) = i \int_0^\infty \Tr\left[ \rho [S^{z \dagger}_{\bm{k}}(t), S^z_{\bm{k}}(0)] \right] e^{i \omega t} \dd t,
\label{eq:chi}
\end{equation}
which, after taking the integral, shares the form
\begin{align}
\chi(\bm{k}, \omega, T) = i \pi \sum_{\lambda, \lambda'} & (p_\lambda - p_{\lambda'}) \left|\!\mel{\lambda}{S^{z \dagger}_{\bm{k}}(0)}{\lambda'} \right|^2 \notag \\
&\times \delta (\omega + E_\lambda - E_{\lambda'}).
\label{eq:chi0}
\end{align}
In reaching the form above, we have adopted the density matrix $\rho = \sum_\lambda p_\lambda \ket{\lambda}\bra{\lambda}$ with $p_\lambda = \frac{1}{Z} e^{-E_\lambda / T}$, and inserted the identity $\mathds{1} = \sum_{\lambda'} \ket{\lambda'}\bra{\lambda'}$.
Comparing \cref{eq:chi0} with \cref{eq:f0} and taking into account
\begin{equation}
\frac{p_\lambda - p_{\lambda'}}{p_\lambda + p_{\lambda'}} = \tanh(\frac{E_{\lambda'} - E_\lambda}{2 T}),
\end{equation}
we arrive at
\begin{equation}
f_\text{Q}(\bm{k}, T) = \frac{4}{\pi} \int_0^\infty \tanh(\frac{\omega}{2 T}) \chi''(\omega, \bm{k}, T)\,\dd \omega,
\label{eq:fq0}
\end{equation}
where $\chi''$ denotes the imaginary (dissipative) part of the dynamical susceptibility.

On the other hand, one can also relate $\chi''$ to the non-symmetrized dynamical spin structure in \cref{eq:sd} through the fluctuation-dissipation theorem. Starting from the integral definition of $\chi$ in \cref{eq:chi}, it is straightforward to identify that
\begin{equation}
\chi''(\bm{k}, \omega, T) = \frac{1}{2} \left( g_1(\bm{k}, \omega, T) - g_2(\bm{k}, \omega, T) \right),
\end{equation}
with
\begin{align}
g_1(\bm{k}, \omega, T) &= \int_{-\infty}^\infty\!\ev*{S^{z \dagger}_{\bm{k}}(t) S^z_{\bm{k}}(0)} e^{i \omega t}\,\dd t, \notag \\
g_2(\bm{k}, \omega, T) &= \int_{-\infty}^\infty\!\ev*{S^z_{\bm{k}}(0) S^{z \dagger}_{\bm{k}}(t)} e^{i \omega t}\,\dd t.
\end{align}
The notation $\ev*{\cdots} = \Tr[\rho \cdots]$ is adopted for a thermal state. Since $g_1 = 2 \pi S_\text{dyn}$, if we can further associate $g_2$ with $g_1$, our goal will be accomplished. It is again more convenient to compare these two quantities after integration. It turns out that
\begin{align}
g_2(\bm{k}, \omega, T) &= \sum_{\lambda, \lambda'} \frac{e^{-E_{\lambda'} / T}}{Z} \left|\!\mel{\lambda}{S^{z \dagger}_{\bm{k}}(0)}{\lambda'} \right|^2 \notag \\
&\phantom{{}= \sum_{\lambda, \lambda'}} \times 2 \pi \delta(\omega + E_\lambda - E_{\lambda'}) \notag \\
&= e^{-\frac{\omega}{T}} g_1(\bm{k}, \omega, T).
\end{align}
Thus we obtain the adapted fluctuation-dissipation theorem,
\begin{equation}
\chi''(\bm{k}, \omega, T) = \pi \left( 1 - e^{-\frac{\omega}{T}} \right) S_\text{dyn}(\bm{k}, \omega, T).
\label{eq:fd}
\end{equation}

Combining \cref{eq:fq0,eq:fd}, the entanglement witness $f_\text{Q}$ is reconstructed from the quantum fluctuations contained in $S_\text{dyn}$, and we reach \cref{eq:fq}:
\begin{equation}
f_\text{Q}(\bm{k}, T) = 4 \int_0^\infty\!\!\tanh(\frac{\omega}{2 T}) \left( 1 - e^{-\frac{\omega}{T}} \right) S_\text{dyn}(\bm{k}, \omega, T)\,\dd \omega.
\end{equation}


%

\end{document}